\documentclass[fleqn,structabstract]{aa}

\usepackage[fleqn]{amsmath}
\usepackage{amssymb}
\usepackage{graphicx}
\usepackage{grffile} 
\usepackage{natbib}
\usepackage{txfonts}
\usepackage{color}
\usepackage{multirow}

\usepackage[]{hyperref}
\hypersetup{
breaklinks = {true},
colorlinks = {false},
pdfpagemode = {None}, 
pdfborder = {0 0 1},
pdftitle = {Galaxy-galaxy(-galaxy) lensing as a sensitive probe of galaxy evolution},
pdfsubject = {study of models of galaxy evolution with galaxy-galaxy(-galaxy) lensing aperture statistics},
pdfauthor = {Hananeh Saghiha, Stefan Hilbert, Peter Schneider, and Patrick Simon},
pdfkeywords = {gravitational lensing: weak -- large-scale structure of the Universe -- galaxies: formation -- galaxies: evolution -- methods: numerical}
}

\everymath{\displaystyle}
\newcommand{\vect}[1]{\boldsymbol{#1}}

\newcommand{\ft}[1]{\tilde{#1}}
\newcommand{\est}[1]{\hat{#1}}
\newcommand{\ev}[1]{\left\langle{#1}\right\rangle}
\newcommand{\bev}[1]{\bigl\langle{#1}\bigr\rangle} 
\newcommand{\abs}[1]{\left\lvert{#1}\right\rvert}

\newcommand{\dd}{\mathrm{d}}
\newcommand{\ee}{\mathrm{e}}
\newcommand{\ii}{\mathrm{i}}

\newcommand{\kms}{\ensuremath{\mathrm{km\,s}^{-1}}}

\newcommand{\kpc}{\ensuremath{\mathrm{kpc}}}
\newcommand{\Mpc}{\ensuremath{\mathrm{Mpc}}}

\newcommand{\Msolar}{\ensuremath{\mathrm{M}_\odot}}

\newcommand{\arcmint}{\ensuremath{\mathrm{arcmin}}}

\newcommand{\clight}{\ensuremath{\mathrm{c}}}

\newcommand{\deltaDirac}{\ensuremath{\delta_\mathrm{D}}}

\newcommand{\Omegab}{\ensuremath{\Omega_\mathrm{b}}}
\newcommand{\Omegam}{\ensuremath{\Omega_\mathrm{m}}}

\newcommand{\vvartheta}{\vect{\vartheta}}
\newcommand{\vell}{\vect{\ell}}

\newcommand{\Map}{M_{\mathrm{ap}}}
\newcommand{\Nap}{\mathcal{N}}

\newcommand{\Ngal}{N_{\mathrm{g}}}
\newcommand{\rhogal}{\rho_{\mathrm{g}}}
\newcommand{\ngal}{n_{\mathrm{g}}}
\newcommand{\pgal}{p_{\mathrm{g}}}
\newcommand{\deltagal}{\delta_{\mathrm{g}}}
\newcommand{\kappagal}{\kappa_{\mathrm{g}}}
\newcommand{\meanrhogal}{\bar{\rho}_{\mathrm{g}}}
\newcommand{\meanngal}{\bar{n}_{\mathrm{g}}}

\newcommand{\estngal}{\est{n}_{\mathrm{g}}}

\newcommand{\rhom}{\rho_{\mathrm{m}}}
\newcommand{\deltam}{\delta_{\mathrm{m}}}
\newcommand{\kappam}{\kappa_{\mathrm{m}}}
\newcommand{\meanrhom}{\bar{\rho}_{\mathrm{m}}}

\newcommand{\Ckgkg}{w_{\mathrm{gg}}}
\newcommand{\Ckgkm}{w_{\mathrm{gm}}}

\newcommand{\Pkgkg}{P_{\mathrm{gg}}}
\newcommand{\Pkgkm}{P_{\mathrm{gm}}}

\newcommand{\BS}{B}
\newcommand{\BSkgkgkm}{\BS_{\mathrm{ggm}}}
\newcommand{\bdgdm}{b_{\mathrm{gm}}}

\begin{document}

\title{Galaxy-galaxy(-galaxy) lensing as a sensitive probe of galaxy evolution}

\author{
Hananeh Saghiha\inst{1}\thanks{Member of the International Max Planck Research School (IMPRS) for Astronomy and Astrophysics at the Universities of Bonn and Cologne.},
Stefan Hilbert\inst{2,3},
Peter Schneider\inst{1}, and
Patrick Simon\inst{1}
}
\authorrunning{Saghiha et al.}

\institute{
$^1$Argelander-Institut f{\"u}r Astronomie, Universit{\"a}t Bonn, Auf dem H{\"u}gel
71, 53121 Bonn, Germany\\
$^2$Kavli Institute of Particle Astrophysics and Cosmology (KIPAC), Stanford
University, 452 Lomita Mall, Stanford, CA 94305, and \\
SLAC National Accelerator Laboratory, 2575 Sand Hill Road, M/S 29, Menlo Park, CA
94025\\
$^3$Max-Planck-Institut f{\"u}r Astrophysik, Karl-Schwarzschild-Stra{\ss}e 1, 85741
Garching, Germany
}

\date{Received / Accepted }

\abstract{
The gravitational lensing effect provides various ways to study the mass environment
of galaxies.
}{
We investigate how galaxy-galaxy(-galaxy) lensing can be used to test models of
galaxy formation and evolution.
}{
We consider two semi-analytic galaxy formation models based on the Millennium Run
N-body simulation: the Durham model by Bower et al. (2006) and the Garching model by
Guo et al. (2011). We generate mock lensing observations for the two models, and
then employ Fast Fourier Transform methods to
compute second- and third-order aperture statistics in the simulated fields for
various galaxy samples.
}{
We find that both models predict qualitatively similar aperture signals, but there
are large quantitative differences. The Durham model predicts larger amplitudes in
general. In both models, red galaxies exhibit stronger aperture signals than blue
galaxies. Using these aperture measurements and assuming a linear deterministic bias
model, we measure relative bias ratios of red and blue galaxy samples. We find that
a linear deterministic bias is insufficient to describe the relative clustering of
model galaxies below ten arcmin angular scales. Dividing galaxies into luminosity
bins, the aperture signals decrease with decreasing luminosity for brighter
galaxies, but increase again for fainter galaxies. This increase is likely an
artifact due to too many faint satellite galaxies in massive group and cluster halos
predicted by the models.
}{
Our study shows that galaxy-galaxy(-galaxy) lensing is a sensitive probe of galaxy
evolution.
}

   
\maketitle

\section{Introduction}
\label{sect:introduction}

Gravitational lensing effects provide versatile tools for
probing the matter distribution in the Universe. Galaxy-galaxy lensing (GGL), for
example, is a
statistical approach using lensing to obtain information on the mass associated with
individual galaxies \citep[see, e.g,][]{2001PhR...340..291B}. This is
achieved by dividing the galaxy population into lenses (foreground)
and sources (background). The images of the sources are sheared due to the
gravitational field of the foreground lenses and their surrounding mass.
The image shearing is usually too small to be detected for individual source-lens
galaxy pairs.
Instead, the lensing effect is measured as a correlation between
the observed image ellipticities and the lens positions. The signal obtained from
averaging over many source-lens pairs can then be related to the average mass
profiles of the lenses.

Since its first detection \citep{1996ApJ...466..623B}, GGL has been measured in many
large galaxy surveys \citep[e.g.,][and references
therein]{2002ApJ...577..604H,2006A&A...455..441K,2006MNRAS.370.1008M,2008A&A...479..655S,2011A&A...534A..14V}.
\citet{2005A&A...432..783S} advanced GGL to galaxy-galaxy-galaxy lensing (G3L) by
introducing third-order correlation functions that involve either configurations
with two background sources and one lens galaxy ($G_{\pm}$), or with two lenses and
one background source ($\cal{G}$). The latter measures the lensing signal around
pairs of lens galaxies in excess of what one obtains by simply adding the average
signals of two individual galaxies, and thus provides a measure of the excess matter
profile about clustered lens galaxy pairs \citep{2012arXiv1202.1927S}. This G3L
signal has been measured in the Red sequence Cluster Survey
\citep[RCS,][]{2005ApJS..157....1G} by \citet{2008A&A...479..655S}, who indeed found
an excess mass about lens pairs with projected separation of $250h^{-1}\,\kpc$. The
GG(G)L correlations can be converted to aperture statistics (which we utilize in
this work), providing a convenient probe of the galaxy-matter power(bi-)spectra at
particular scales.

The galaxy-mass correlation as seen by weak lensing can also be studied
theoretically by combining dark matter simulations with semi-analytic models (SAM)
of galaxy evolution \citep{1991ApJ...379...52W, 1999MNRAS.303..188K,
2001MNRAS.328..726S}. In this approach, the dark matter halos of an $N$-body
simulation of cosmic structure formation are populated with galaxies. The properties
of the galaxies are calculated by combining information on the halo merger trees of
the underlying dark matter simulation with an analytic model of the gas physics in
galaxies. The physical processes considered include gas cooling, star formation,
metal enrichment, and feedback due to supernovae and active galactic nuclei. Using
ray-tracing \citep[e.g.][]{2009A&A...499...31H}, one can then simulate lensing
observations of the resulting galaxy distribution.

This paper provides a study of the second- and third-order galaxy-mass correlations
in semi-analytic galaxy formation models as probed by lensing via aperture
statistics \citep{1996MNRAS.283..837S,1998MNRAS.296..873S}. We consider two models
based on the Millennium Run \citep{2005Natur.435..629S}: the \emph{Durham} model by
\cite{2006MNRAS.370..645B} and the \emph{Garching} model by
\citet{2011MNRAS.413..101G}. We find that the predicted second- and third-order
lensing signals differ between galaxies of different color and magnitude, but also
between the different galaxy models. The differences between the models can be
traced back to, among other things, different treatments of the satellite galaxy
evolution. This illustrates that galaxy-galaxy\-\mbox{(-galaxy)} lensing can be a
sensitive probe of galaxy evolution.

The outline of the paper is as follows: Sect. \ref{sect:theory} provides a brief
account of gravitational lensing, aperture statistics, and their relation to
correlation functions. 
Our lensing simulations and the method we use to measure aperture statistics (a fast
method based on Fast Fourier Transforms) are described in Sect. \ref{sect:methods}.
The results of these measurements for different subsets of galaxies, defined by
redshift, luminosity or color, are presented in Sect. \ref{sect:results}. 
The main part of the paper concludes with a summary and discussion in Sect.
\ref{sect:discussion}. In the appendix, we briefly discuss shot-noise corrections
for the aperture statistics.

\section{Theory}
\label{sect:theory}

\subsection{Gravitational lensing basics}
\label{sect:lensing_preliminaries}

The matter density inhomogeneities can be quantified by the dimensionless density
contrast
\begin{equation}
\label{eq:deltam}
 \deltam(\vect{x},\chi) = \frac{ \rhom(\vect{x},\chi) -
\meanrhom(\chi)}{\meanrhom(\chi)},
\end{equation}
where $\rhom(\vect{x},\chi)$ is the spatial matter density at comoving transverse
position $\vect{x}$ and comoving radial distance $\chi$, and $\meanrhom(\chi)$
denotes the mean density at that distance.
To lowest order, the convergence $\kappam$ for sources at comoving distance $\chi$
is related to the matter density contrast $\deltam$ by the projection along the
line-of-sight by \citep[e.g.][]{schneider2006gravitational}
\begin{equation}
\label{eq:bornkappa}
\kappam(\vvartheta ,\chi) = \frac{3 H_0^2 \Omegam}{2 \clight^2} \!\!
\int_{0}^{\chi}\!\!\! \dd\chi'\, \frac{f_K(\chi - \chi') f_K(\chi')}{f_K(\chi)} 
\frac{\deltam(f_K(\chi')\vvartheta , \chi')}{a(\chi')}, 
\end{equation}
where $\kappam$ describes the dimensionless projected matter density,
$H_0$ denotes the Hubble constant, $\Omegam$ the mean matter density parameter,
$\clight$ the speed of light, $f_K(\chi)$ the comoving angular diameter distance,
and $a(\chi) = 1 / (1 + z(\chi))$  the scale factor at redshift $z(\chi)$.

For a distribution of sources with probability density $p_{\mathrm{s}}(\chi)$, the
effective convergence is given by 
\begin{align}
\kappam (\vvartheta) &= \int \dd\chi \, p_{\mathrm{s}}(\chi)\,\kappam(\vvartheta ,\chi)
\nonumber\\ 
\label{eq:kappagfunction}
& =\int \dd \chi \, g(\chi) \deltam(f_K(\chi)\vvartheta, \chi)
\quad\text{with}
\\
g(\chi) &=
  \frac{3H_{0}^{2}\Omegam}{2 \clight^{2}} \frac{ f_K(\chi)}{a(\chi)}
   \int_{\chi}^{\infty} \dd \chi' \, p_{\chi}(\chi') \frac{f_K(\chi'-\chi)}{f_K(\chi')}
.
\end{align}

Similar to the definition of the dimensionless matter density contrast $\deltam$,
one can define the number density contrast $\deltagal$ of the lens galaxies as
\begin{equation}
 \deltagal(\vect{x},\chi) = \frac{ \rhogal(\vect{x},\chi) -
\meanrhogal(\chi)}{\meanrhogal(\chi)},
\label{eq:deltag}
\end{equation}
where $\rhogal(\vect{x},\chi) $ is the number density of the lens galaxies at
comoving transverse position $\vect{x}$ and distance $\chi$, and $\meanrhogal$ is
mean number density of lens galaxies at comoving distance $\chi$. Using the
projected number density
\begin{equation}
        \ngal(\vvartheta) = \int \dd \chi f_K^2(\chi) \, \rhogal(f_K(\chi)\vvartheta , \chi)
\end{equation}
and the mean projected number density 
\begin{equation}
        \meanngal = \int \dd \chi f_K^2(\chi) \, \meanrhogal(\chi),
\end{equation}
the projected number density contrast for lens galaxies with distance distribution
$\pgal(\chi) = \meanngal^{-1} f_K^2(\chi) \, \meanrhogal(\chi)$
can be computed by
\begin{equation}
  \label{eq:1}
\kappagal(\vvartheta)
= \frac{\ngal(\vvartheta) - \meanngal}{\meanngal}
 = \int \dd \chi \, \pgal (\chi) \deltagal(f_K(\chi)\vvartheta , \chi)
.
\end{equation}

\subsection{Aperture Statistics}
\label{sect:aperture}

Aperture statistics was originally introduced as a way to quantify the surface mass
density that is unaffected by the mass sheet degeneracy \citep{1996MNRAS.283..837S}.
The aperture mass is defined as a convolution,
\begin{equation}
   \Map(\vvartheta;\theta) = \int \dd^{2}\vvartheta' \, U_\theta(\vert \vvartheta -
\vvartheta' \vert)\, \kappam(\vvartheta'),
  \label{eq:map}
\end{equation}
of the convergence $\kappa$ and an axi-symmetric filter $U_\theta(\lvert \vvartheta
\rvert)$ whose size is given by the scale $\theta$, and that is compensated, i.e.
\begin{equation}
 \int\dd \vartheta\, \vartheta\, U_{\theta}(\vartheta) = 0.
\end{equation} 
In this work, we use the filter introduced by \cite{2002ApJ...568...20C},
\begin{align}
\label{eq:ux}
U_{\theta}(\vartheta) &= \frac{1}{2\pi \theta^{2}} \left [ 1 -
\frac{\vartheta^{2}}{2 \theta^2} \right ] \exp \left
(\frac{-\vartheta^{2}}{2\theta^{2}} \right)
= \frac{1}{\theta^{2}}\,u \left (\frac{\vartheta}{\theta} \right )
  \quad \text{with} \\
  u(x)  &= \frac{1}{2\pi} \left [ 1 - \frac{x^{2}}{2} \right ] \exp \left
(\frac{-x^{2}}{2} \right)
.
\end{align}
Its (2-D) Fourier transform has a simple analytical form
\begin{align}
\ft{U}_{\theta}(\ell) &= \int \dd^{2} \vvartheta \, U_{\theta} (\vert \vvartheta
\vert) \,\ee^{\ii \vell\cdot \vvartheta}
 = \frac{\theta^2\ell^{2}}{2}\, \mathrm{e}^{-\frac{1}{2} \theta^2 \ell^{2} }
 = \ft{u}(\theta \ell)
 \quad\text{with}\\
 \ft{u}(k)  &=  \frac{k^{2}}{2} \exp \left( \frac{-k^{2}}{2} \right)
.
\end{align}
The filter falls off exponentially for $\vartheta \gg \theta$. This makes the
support of the filter
finite in practice.

In analogy to the aperture mass $\Map$, one can define the aperture number count
\begin{equation}
  \Nap(\vvartheta; \theta)= \int \dd^{2} \vvartheta' \, U_{\theta} (\vert \vvartheta
- \vvartheta' \vert) \, \kappa_{\mathrm{g}} ( \vvartheta').
  \label{eq:N}
\end{equation}
The aperture number count dispersion is related to the angular two-point correlation
function
\begin{equation}
 \Ckgkg (\lvert \vvartheta_{2} - \vvartheta_{1} \rvert) =  \ev{\kappagal (
\vvartheta_{1} ) \kappagal ( \vvartheta_{2} )},
\end{equation}
and its Fourier transform, the angular power spectrum $\Pkgkg(\ell)$ of the lens
galaxies, through
\begin{equation}
 \label{eq:apr_num_dis}
\begin{split}
&
 \bev{\Nap^2} (\theta)
 \equiv
 \ev{\Nap(\vvartheta; \theta) \Nap(\vvartheta; \theta)}
 \\&\quad=
  \int \dd^{2} \vvartheta_1 \, U_{\theta} (| \vvartheta_1|) \, \int \dd^{2}
\vvartheta_2 \, U_{\theta} ( |\vvartheta_2|) \,
 \Ckgkg(\lvert \vvartheta_{2} - \vvartheta_{1} \rvert)
 \\&\quad=
 \int_0^{\infty}\frac{\ell \dd \ell}{2\pi} \ft{u}^2(\theta \ell) \Pkgkg(\ell)
.
\end{split}
\end{equation} 
The function $\ft{u}(\ell \theta)$ features a sharp peak at $\ell \theta =
\sqrt{2}$. Thus, $\bev{\Nap^2} (\theta)$ provides a measurement of the corresponding
power spectrum $\Pkgkg(\ell)$ at wave numbers $\ell \sim 1/\theta$.

Within the halo model framework of cosmic structure
\citep[e.g.][]{2002PhR...372....1C}, $\bev{\Nap^2}(\theta)$ on small scales $\theta$
probes the distribution of the lens galaxies within individual dark matter halos. On
large scales, $\bev{\Nap^2}$ provides a probe of the clustering of the host halos of
the lens galaxies.

Correlating $\Map(\theta)$ with $\Nap(\theta)$ yields 
\begin{equation}
\begin{split}
 \label{eq:nm}
&
\ev{\Nap \Map}(\theta)
\equiv
   \ev{\Nap(\vvartheta;\theta) \Map(\vvartheta; \theta)}
\\&\quad=
  \int \dd^{2} \vvartheta_{1} \, U_{\theta} ( |\vvartheta_{1}|) \, \int \dd^{2}
\vvartheta_{2} \, U_{\theta} ( |\vvartheta_{2}|) 
  \Ckgkm(\lvert \vvartheta_{2} - \vvartheta_{1} \rvert)
\\&\quad=
  \int_0^{\infty}\frac{\ell \dd \ell}{2\pi} \ft{u}^2(\theta \ell) \Pkgkm(\ell)
,
\end{split}
\end{equation} 
with $\Ckgkm(\lvert \vvartheta_{2} - \vvartheta_{1} \rvert) =  \bev{\kappagal (
\vvartheta_{1} ) \kappam ( \vvartheta_{2} )}$, whose Fourier transform is the
cross-power spectrum of galaxies and convergence $\Pkgkm$. The galaxy-galaxy lensing
aperture statistics $\bev{\Nap \Map}$ probes the average matter profiles around lens
galaxies.

A third-order aperture correlator (\citealt{2005A&A...432..783S}) is obtained by 
\begin{equation}
\label{eq:mmn}
\begin{split} 
&
\ev{\Nap^2 \Map} (\theta) 
\equiv
  \ev{\Nap(\vvartheta;\theta) \Nap(\vvartheta;\theta) \Map(\vvartheta;\theta )}
\\&\quad =
  \int \dd^{2} \vvartheta_{1} \, U_{\theta} ( |\vvartheta_{1}|) \, \int \dd^{2}
\vvartheta_{2} \, U_{\theta} ( |\vvartheta_{2}|) \, \int \dd^{2} \vvartheta_{3} \,
U_{\theta} ( |\vvartheta_{3}|) \\ 
& \quad \quad \times
   \left\langle \kappa_{\mathrm{g}} ( \boldsymbol {\vartheta}_{1} )
\kappa_{\mathrm{g}} ( \boldsymbol {\vartheta}_{2} ) \kappa ( \boldsymbol
{\vartheta}_{3} ) \right\rangle \\
&\quad=
  \int\frac{\dd^{2}\vell_{1}}{(2\pi)^{2}}\int\frac{\dd^{2}\vell_{2}}{(2\pi)^{2}} \,
\ft{u}(\theta\abs{\vell_{1}}) \, \ft{u}(\theta\abs{\vell_{2}}) \, 
\ft{u}(\theta\abs{\vell_{1} + \vell_{2}})  \\
&\quad \quad \times
   \BSkgkgkm (\vell_{1} , \vell_{2}, -\vell_{1} - \vell_{2} )
\end{split}
\end{equation}
where the last line contains the angular bispectrum of the projected quantities
\citep[][]{2005A&A...432..783S},
\begin{equation}
    \left\langle \ft{\kappa}_{1} (\vell_{1})\ft{\kappa}_{2}
(\vell_{2})\ft{\kappa}_{3}(\vell_{3}) \right\rangle 
  = (2\pi)^{2}\delta_{\mathrm{D}} (\vell_{1} + \vell_{2} + \vell_{3})
\BS_{123}(\vell_{1} , \vell_{2} , \vell_{3}).
  \label{eq:relation_nnm_bsp}
\end{equation}
On small scales, $\bev{\Nap^2 \Map}$ can teach us about the average mass
distribution of halos hosting two lens galaxies. On larger scales, $\bev{\Nap^2
\Map}$ also provides information on the higher-order clustering of the host halos.

\subsection{Relative galaxy bias}
\label{sec:galaxy_bias}

Clusters and galaxies are biased tracers of the matter distribution
\citep[][]{1984ApJ...284L...9K,1986ApJ...304...15B,1996MNRAS.282.1096M}. In the
simplest conceivable non-trivial bias model, the bias can be expressed as a linear
deterministic relation between the galaxy density contrast and the matter density
contrast,
\begin{equation}
  \deltagal (\vect{x}, \chi) = \bdgdm \deltam (\vect{x}, \chi),
\end{equation}
with a bias factor $\bdgdm$ that does not depend on time or scale, but only on the
galaxy sample in question.

A more realistic assumption is that the galaxy bias is stochastic and depends on the
time and spatial scale. At the two-point level, the bias may then be quantified by a
scale-dependent bias factor and correlation factor.
A description of the relation between the galaxy and matter densities at the
three-point level or higher requires additional, higher-order bias and correlation
parameters.

Aperture statistics can be used to constrain the galaxy bias
\citep[][]{1998ApJ...498...43S,1998A&A...334....1V,2002ApJ...577..604H,2005A&A...432..783S,2007A&A...461..861S,2012arXiv1202.6491J}.
For example, assuming a linear deterministic bias, $\bev{\Nap^2} \propto b^2$,
$\bev{\Nap \Map} \propto b$, and $\bev{\Nap^2 \Map} \propto b^2$. For two lens
galaxy samples with identical redshift distributions, but different bias parameters
$b_1$ and $b_2$, one can then determine the relative bias $b_1 / b_2$  from the
aperture statistics  $\bev{\Nap^2_{1/2}}$, $\bev{\Nap_{1/2} \Map}$, and
$\bev{\Nap^2_{1/2} \Map}$ of the two lens samples by using any of
\begin{align}
\frac{b_1}{b_2} &=
 \sqrt{\frac{\bev{\Nap^2_1}(\theta)}{\bev{\Nap^2_2}(\theta)}}
= \frac{\bev{\Nap_1 \Map}(\theta)}{\bev{\Nap_2 \Map}(\theta)}
= \sqrt{\frac{\bev{\Nap_1^2 \Map}(\theta)}{\bev{\Nap_2^2 \Map}(\theta)}}
.
\end{align}

If the measured ratios of the aperture statistics change with scale $\theta$, one
can extend the idea to a measurement of a scale-dependent bias. 
For lens galaxy samples with narrow redshift distributions and deterministic bias,
the above ratios still agree (roughly) when compared on the same scales. If the
galaxy bias is stochastic or non-linear \citep{1999ApJ...520...24D}, however, the ratios from the different statistics
disagree even if measured on the same scale. In that case, the second- and
third-order aperture statistics each contain valuable independent information on the
second- and third-order bias of the galaxies \citep{2005A&A...432..783S}.

\section{Methods}
\label{sect:methods}

\subsection{Lensing simulations}
\label{sect:simulations}

For our analysis we use the data obtained by ray-tracing through the Millennium Run
\citep{2005Natur.435..629S}. The Millennium Run (MR) is a large N-body simulation of
structure formation in a flat $\Lambda$CDM universe with matter density $\Omegam =
0.25$, baryon density $\Omegab = 0.045$, dark-energy density $\Omega_\Lambda =
0.75$, a Hubble constant $H_0 = h 100 \kms \Mpc^{-1}$ with $h = 0.73$, and with a
power spectrum normalization $\sigma_{8} = 0.9$. It follows the evolution of
$N_\mathrm{p}\sim 10^{10}$ dark matter particles with mass $m_{\mathrm{p}} = 8.6
\times 10^{8} h^{-1}\,\Msolar$ in a cubic region of comoving side length
$500h^{-1}\,\Mpc$ from redshift $z = 127$ to the present.

The simulation volume of the MR is large enough to include massive rare objects, yet
with sufficiently high spatial and mass resolution to resolve dark matter halos of
galaxies. This allows the construction of merger trees of dark matter halos and
subhalos within them. These merger trees have been used in various semi-analytic
galaxy formation models to calculate the properties of galaxies in the simulation.
Here we consider the \emph{Durham model} by  \citet{2006MNRAS.370..645B} and the
\emph{Garching model} by \citet{2011MNRAS.413..101G}.\footnote{
We also considered an earlier incarnation of the Garching model by
\citet{2007MNRAS.375....2D}. However, the differences between the results from two
Garching models are minor. Thus, we concentrate the discussion on the models by
\citet{2006MNRAS.370..645B} and \citet{2011MNRAS.413..101G}.
} 
Both models have similar treatments of, e.g., gas cooling and star formation, but
differ in various details (see the original papers for a full description). 
The models have been adjusted to be consistent with a large number of observations,
in particular the luminosities, stellar masses, morphologies, gas contents and
correlations of galaxies at low redshift, but they have not been tuned to match
galaxy properties at higher redshift.
We make use of the public Millennium Simulations
Database\footnote{\texttt{http://www.mpa-garching.mpg.de/millennium/}}
\citep[][]{2006astro.ph..8019L} to obtain the properties of the galaxies predicted
by the two models.

We employ the multiple-lens-plane ray-tracing algorithm described in
\citet[][]{2009A&A...499...31H} to calculate the light propagation through the
matter in the MR. We generate 64 simulated $4 \times 4\,\mathrm{deg}^2$ fields of
view. For each field, we calculate the convergence to sources at a number of
redshifts on a regular mesh of $4096^2$ pixels, as well as the apparent sky
positions, redshifts, and magnitudes of the model galaxies from the Garching and
Durham models. The galaxy properties are then used to select various subsamples of
the full mock galaxy catalogs as lens populations for the GG(G)L.

\subsection{Computing aperture statistics}
\label{sect:computing_aperture_statistics}

We introduce a fast method to perform aperture statistics measurements on the 64
simulated fields. The statistic used here is built on two main components:
(i) the pixelized convergence field $\kappam(\vvartheta)$ of the source galaxies on
square meshes of $4096^2$ pixels, and (ii) the pixelized lens galaxy number density
fields on meshes with the same geometry. The convergence fields are obtained
directly from the ray-tracing algorithm. The galaxy density fields are obtained by
projecting the apparent position of the lens galaxies in the fields and counting the
number of galaxies in each pixel. Finally, dividing by the mean number density of
lens galaxies across all 64 fields results in the galaxy number density contrast
$\kappagal(\vvartheta)$. 

We calculate the aperture statistics $\Map(\vec\vartheta;\theta)$ and $\mathcal
{N}(\vec\vartheta;\theta)$ from $\kappam$ and $\kappagal$ on a grid by exploiting
the convolution theorem, using Fast Fourier Transforms \citep[FFT, in particular the
FFTW library by][]{FrigoJo05} to carry out the convolution in Eqs. \eqref{eq:map}
and \eqref{eq:N}.
To measure $\Map$ (or $\mathcal {N}$), we calculate the Fourier transforms
of $ \kappam$ (or $\kappagal$) and $U_\theta(\vert \boldsymbol{\vartheta} \vert)$.
We then multiply the results in the Fourier space. Finally an inverse Fourier
transformation gives $\Map$ (or $\mathcal {N}$). The number of grid points in the
field is finite (4096$\times$4096 pixels). Therefore, a ``Discrete Fourier
Transform'' is performed by using the Fast Fourier Transform (FFT) algorithm
which reduces an $ O(N^{2})$ process to $O(N\,\log_{2}N)$, with $N$ being the number
of points being Fourier transformed. Hence, FFT reduces the computation time
immensely.

Since $\Map(\vec\vartheta;\theta)$ and $\mathcal
{N}(\vec\vartheta;\theta)$ fields are not periodic, we exclude points closer than
$4\theta$ to the field edges from the subsequent analysis. On the remaining points,
we then calculate 
$\Nap^2(\vvartheta; \theta)$, $\Nap(\vvartheta; \theta) \Map(\vvartheta; \theta)$,
and  $\Nap^2(\vvartheta; \theta) \Map(\vvartheta; \theta)$, 
and estimate $\bev{\Nap^{2}}(\theta)$, $\bev{\Nap\Map}(\theta)$, and
$\bev{\Nap^{2}\Map}(\theta)$ from these products by spatial averaging. We correct
the estimates involving $\Nap^2$ for shot noise as described in Appendix
\ref{sec:shot_noise_correction}.

The G3L statistics can also be calculated from the shear field corresponding to
$\kappam$ and the positions of the lens galaxies. In particular,
\citet{2005A&A...432..783S} and \citet{2008A&A...479..655S} showed that
$\bev{\Nap^{2}\Map}$ can be obtained as an integral over a three-point correlation
function (3PCF). In order to check our procedure, we use 32 randomly selected
simulated fields and calculate $\bev{\Nap^{2}\Map}$ also with the latter method, by
first calculating this 3PCF with the help of a tree code. We note that, while this
tree method is more flexible than the simple FFT-based method described above, in
particular with regard to field boundaries and gaps, it is also considerably slower.

For some of the individual fields, we find fairly large discrepancies between the
results from the FFT and the tree method -- in particular for fields with a large
matter overdensity near the field boundaries. We can attribute these discrepancies
to the different ways in which the three-point information is weighted in the two
approaches. For example, a triplet of points near the boundary of the field enters
the statistics in the tree method with the same weight as a similar triplet near the
field center. In contrast, the FFT method, by excluding the stripe at the field
boundary, assigns zero weight to such a triple. Hence, the results on individual
fields can be quite different.

\begin{figure}
\centerline{\includegraphics[width=92mm]{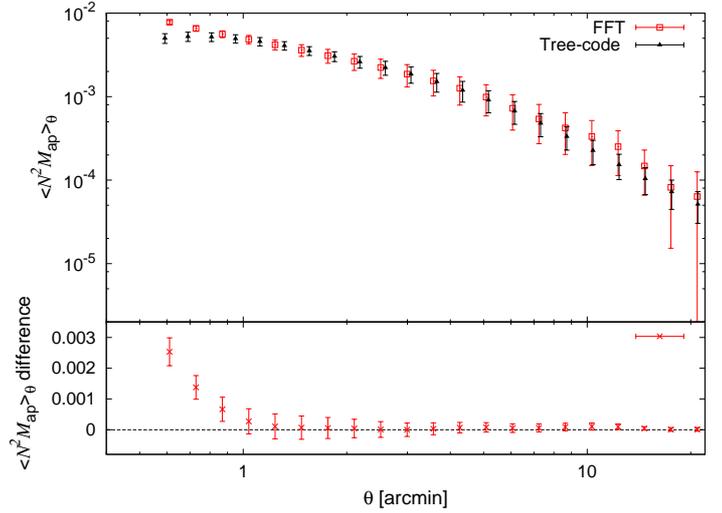}}
\caption{
Upper panel: Aperture statistics $\bev{\Nap^{2}\Map}(\theta)$ as a function
of filter scale $\theta$ measured in the Garching model. The FFT method (squares)
and the tree method (triangles) are compared for lenses at redshift $z = 0.17$ with
m$_{\mathrm{r}} \leq 22.5$, stellar masses $M_{\star} \geq 10^{9}h^{-1}\,\Msolar$
and convergence field of sources at redshift $z = 0.99$. Error bars indicate the
standard deviation of $\bev{\Nap^{2}\Map}(\theta)$ for aperture radius $\theta$
estimated across 32 fields. Lower panel: Average difference signal between the FFT
method and the tree method. Again the error bars show the standard deviation of the
mean (field variance of difference signal divided by $\sqrt{31}$).}
\label{fig:NNM.onplane56}
\end{figure}

Both methods are consistent, however, when averaging the results over many fields. Randomly selecting 32 simulated fields, we measure
$\bev{\Nap^{2}\Map}(\theta)$ using the FFT method and the tree method. In the upper
panel of Fig. \ref{fig:NNM.onplane56}, the outcomes of the two methods are
compared, showing good agreement between the results. The error bars, indicating the
statistical error on the signal, tend to be smaller for the tree method than for the
FFT method (for apertures larger than 2 \,\arcmint), since the tree method makes better use of the fields' area. For example,
for apertures larger than 20 \,\arcmint, more than half of the field is not included in
the FFT measurement. Consequently, the difference in scatter becomes more prominent
on larger scales. The lower panel in Fig. \ref{fig:NNM.onplane56} shows the
field-by-field difference signal averaged over all fields. The difference between
the methods is consistent with zero for $\theta \geq 1\,\arcmint$, but deviates from
zero for  $\theta < 1\,\arcmint$. This is due to a systematic underestimation of the
signal in the tree method on small scales \citep{2008A&A...479..655S}.


\section{Results}
\label{sect:results}

\subsection{Main lens samples}
\label{sect:results_main_lens_sample}

\begin{figure}
\centerline{\includegraphics[width=92mm]{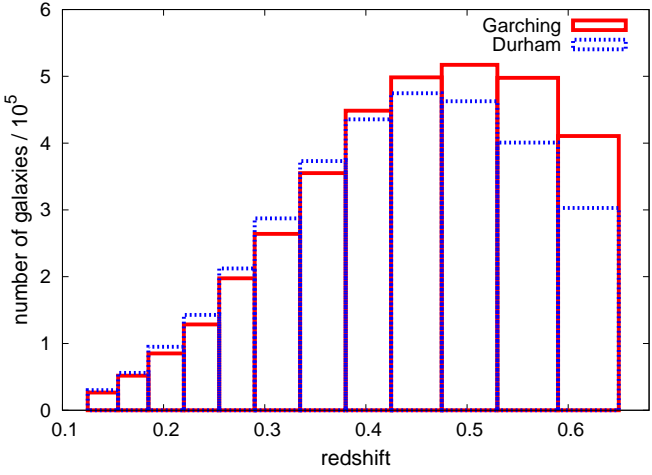}}
\caption{
Redshift distribution of galaxies in the main lens samples (i.e. galaxies with
redshifts $0.14 \leq z \leq 0.62$, observer-frame $r$-band apparent magnitude $m_r
\leq 22.5$, and stellar masses $M_{\star} \geq 10^{9}h^{-1} \Msolar$) in the
Garching and Durham model.}
\label{fig:redshift_distribution_main_sample}
\end{figure}

In this section, we present results for the second- and third-order aperture
cross-correlations and aperture number count dispersion for the Durham and the
Garching model in the 64 simulated fields created from the Millennium Run. For
simplicity, the background population is chosen to be located at $z = 0.99$.
Unless stated otherwise, lens galaxies are selected to have redshifts $0.14 \leq z
\leq 0.62$, observer-frame $r$-band apparent magnitude $m_r \leq 22.5$, and stellar
masses $M_{\star} \geq 10^{9}h^{-1}\,\Msolar$. This yields $8.5\times10^6$ lens
galaxies in the Durham model and  $8.7\times10^6$ galaxies in the Garching model.
The resulting redshift distributions for the lens populations are shown in
Fig.~\ref{fig:redshift_distribution_main_sample}.

\begin{figure}
\centerline{\includegraphics[width=80mm]{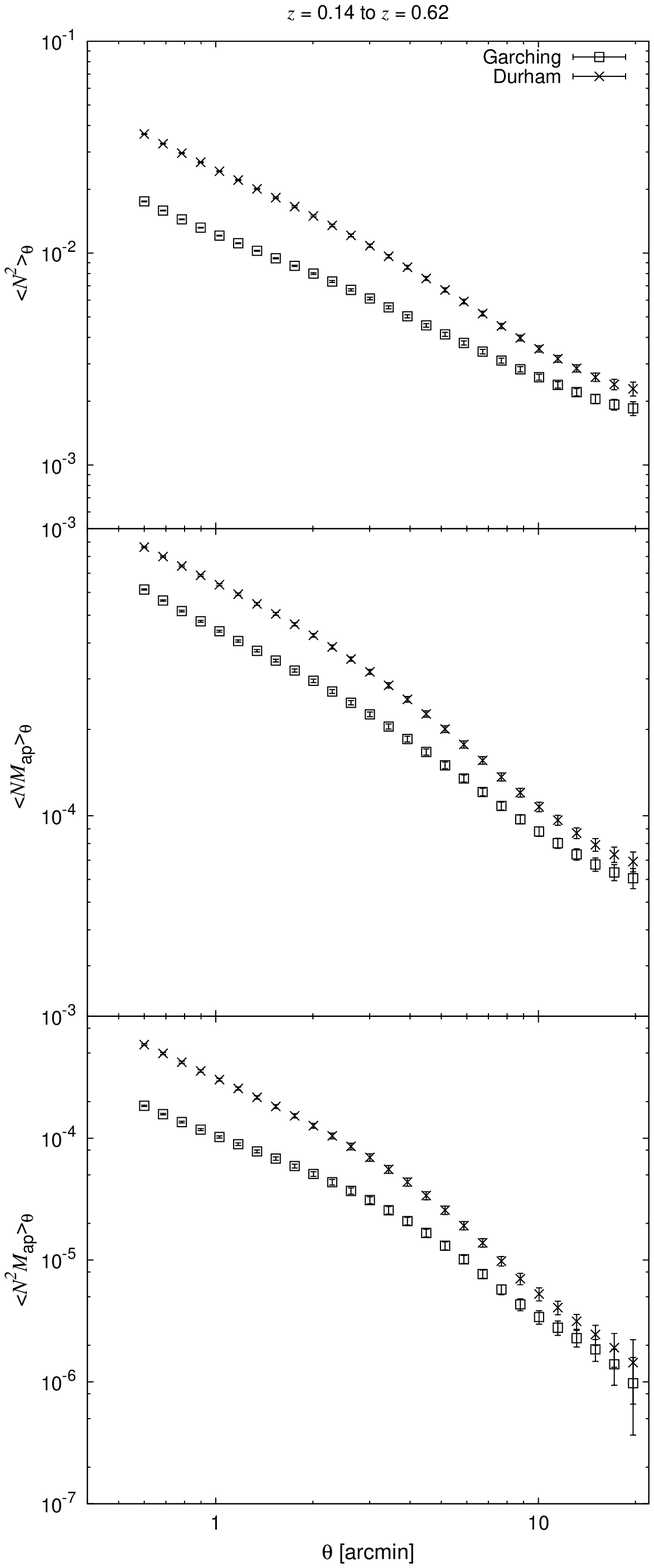}}
\caption{Aperture number count dispersion (top panel), $ \bev{\Nap\Map}$ (middle
panel) and $ \bev{\Nap^{2}\Map}$ (bottom panel) measurements in the Garching model
and Durham model.}
\label{fig:guo_durham_NN_NM_NNM_paper}
\end{figure}

For this sample of galaxies, the aperture number count dispersion $\bev{\Nap^{2}}$
as a function of aperture radius is shown in the top panel of Fig.
\ref{fig:guo_durham_NN_NM_NNM_paper}. The galaxy models clearly differ in the
predicted dispersion: the Durham model predicts an up to two times larger amplitude
than the Garching model. A similar difference has been observed for the angular
galaxy correlation function by \citet{2009MNRAS.400.1527K}, who attribute the
discrepancy to too many bright satellites in the Durham model. However, as will be
discussed below, the Garching model also appears to suffer from problems with the
modeling of the satellite population.

The predictions for $\bev{\Nap\Map}$, shown in the middle panel of Fig.
\ref{fig:guo_durham_NN_NM_NNM_paper}, exhibit fairly large differences between the
models, too. The higher values of $\bev{\Nap\Map}$ in the Durham model, especially
for smaller angular scales, imply more massive lens halos on average compared to the
Garching model. The larger halo masses may also explain the higher clustering
amplitude seen in $\bev{\Nap^{2}}$. More massive halos host larger concentrations of
galaxies and are themselves more clustered, which increases the correlation of the
hosted galaxies on small and large scales. Another consequence is a larger
third-order signal $\bev{\Nap^{2} \Map}$, which is confirmed by the bottom panel of
Fig. \ref{fig:guo_durham_NN_NM_NNM_paper}.

\subsection{Color-selected samples}

For a further analysis, we divide the main lens galaxy samples into groups selected
by color. From observations, the color distribution of galaxies is well
characterized by a bimodal function \citep{2001AJ....122.1861S}. At low redshifts,
this can be approximated by the sum of two Gaussian functions representing red and
blue subpopulations of galaxies on the red and blue side of the color distribution,
respectively.

\begin{figure}
\centerline{\includegraphics[width=90mm]{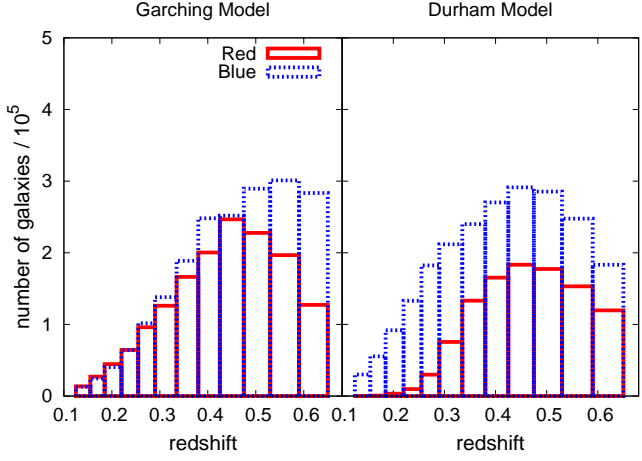}}
\caption{Number of red and blue galaxies in the Garching model and the Durham model.
Red galaxies are selected to have $u-r > 2.2$ and blue galaxies are to have $u-r
\leqslant 2.2$.}
\label{fig:durham_garching_zdistribution_46-57.2.2}
\end{figure}

Following observations \citep[e.g.][]{2006MNRAS.370.1008M}, we split the main lens
galaxy samples at observer-frame color $u-r$ = 2.2 to obtain subsamples of red and
blue lens galaxies. For the Durham model, we obtain $2.4 \times$10$^{6}$ red and
$6.1\times$10$^{6}$ blue galaxies compared to 
 $4\times$10$^{6}$ red and $4.7\times$10$^{6}$ blue galaxies in the Garching model.
The redshift distributions of the color subsamples are shown in
Fig.\ref{fig:durham_garching_zdistribution_46-57.2.2}. The histograms show that
the relative numbers of red and blue galaxies in each redshift bin differ
significantly between the models.

\begin{figure*}
\centerline{\includegraphics[width=160mm]{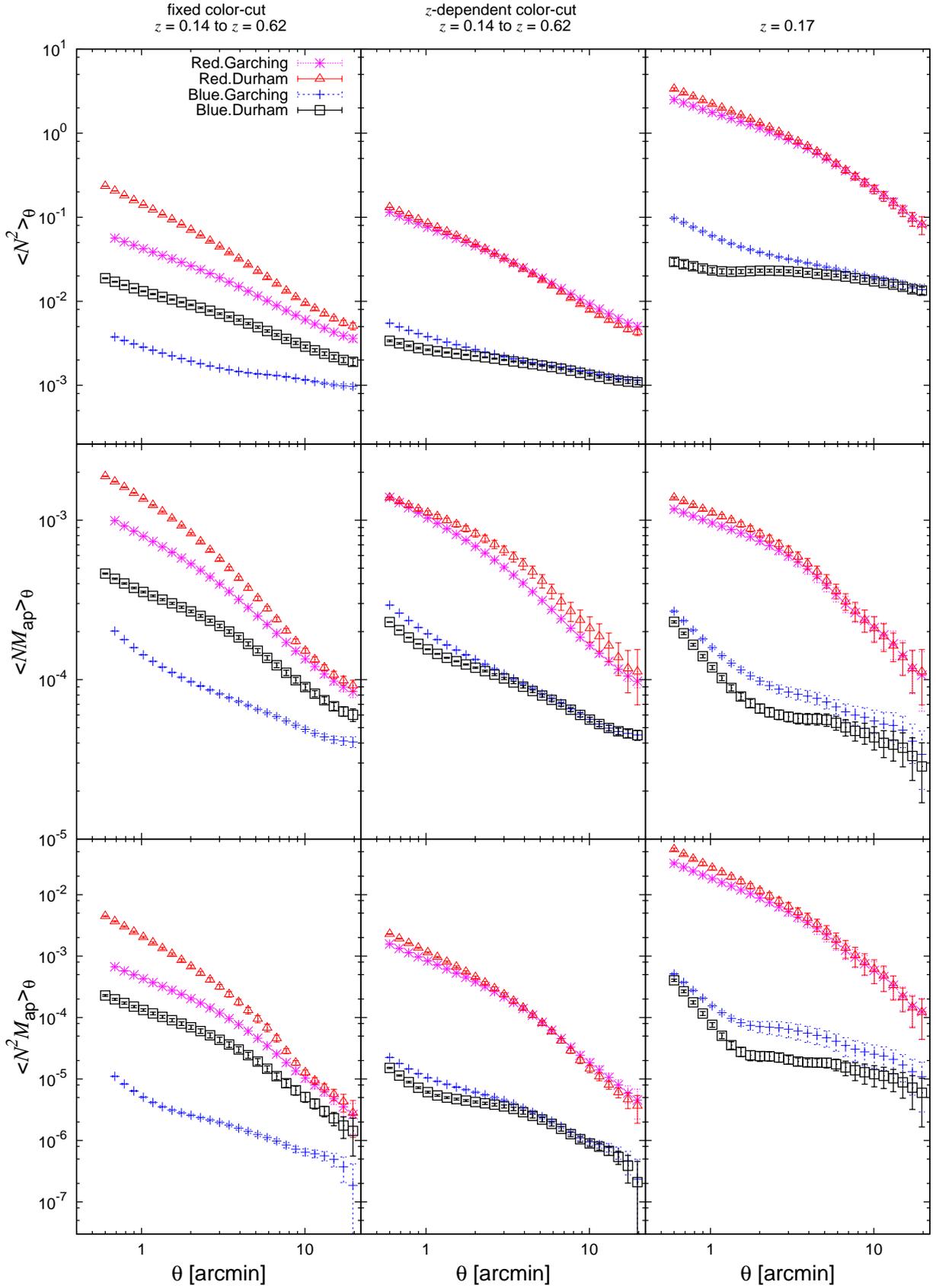}}
\caption{Aperture statistics for samples of red and blue galaxies in the Garching
and Durham models. The left column shows the results for a fixed color-cut at $u -
r$ = 2.2 for galaxies with redshift between $z = 0.14$ and $z = 0.62$. The middle
column displays the signals for galaxies between $z = 0.14$ and $z = 0.62$ separated
using a redshift-dependent color cut. In the right column, galaxies are restricted to come from a single snapshot at
redshift $z = 0.17$; accordingly error bars are larger.}
\label{fig:guo_durham_NN_NM_NNM_paper_red_blue}
\end{figure*}

The aperture statistics for these red and blue galaxy populations in both models are
shown in the left panels of Fig. \ref{fig:guo_durham_NN_NM_NNM_paper_red_blue}. In
both models, red galaxies show higher signals than the blue galaxies. This trend is
not surprising, since galaxies of different types follow different distribution
patterns and clustering properties. Red galaxies are expected to be found mainly in
groups and clusters associated with strong clustering and large halo masses, whereas
blue galaxies are mostly field galaxies with smaller halos and lower clustering. The
plot also shows that galaxies in the two models show different clustering statistics
in both red and blue populations. This may be a result of selecting different
objects in the models.

\begin{figure}
\centerline{\includegraphics[width=90mm]{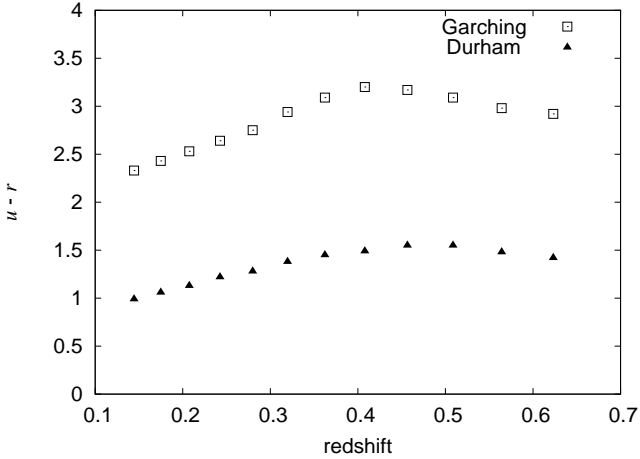}}
\caption{
The $u -r$ color-cut at each redshift in the Garching and Durham models.}
\label{fig:bimodal}
\end{figure}

\begin{figure}
\centerline{\includegraphics[width=90mm]{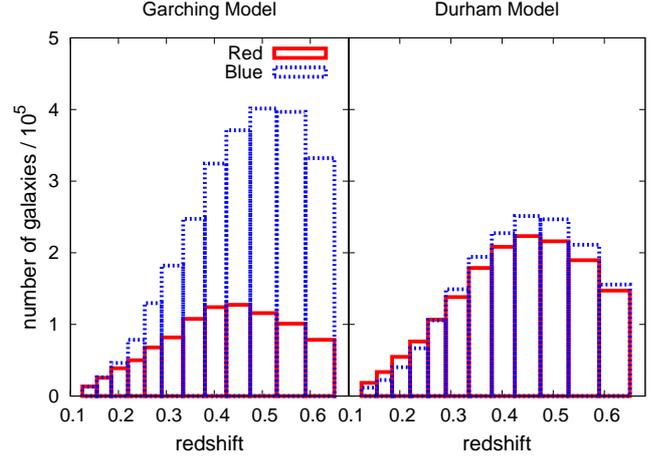}}
\caption{
Number of red and blue galaxies counted in the Garching model and the Durham model.
Galaxies at each redshift are selected according to the color-cut in Fig.
\ref{fig:bimodal}.}
\label{fig:durham_garching_zdistribution_46-57.manual}
\end{figure}

Studying the previous version of the Garching model based on the
\citet{2007MNRAS.375....2D}, \cite{2012MNRAS.420.3303B} pointed out that the
distributions of the observer-frame $u - r$ colors are very different in the
Garching model and the Durham model. In particular, galaxies appear redder in the
Garching model than in the Durham model. We thus consider another way of dividing
the galaxies into red and blue samples. We identify the minima of the bimodal
distributions at each redshift. The positions of the minima are plotted in Fig.
\ref{fig:bimodal}, clearly showing a large difference between the models in their
color distribution. We then use the minima to separate red and blue galaxies.

Fig. \ref{fig:durham_garching_zdistribution_46-57.manual} shows the resulting
redshift distributions of the red and blue subsamples. There are $4.2
\times$10$^{6}$ red and $4.3\times$10$^{6}$ blue galaxies in the Durham model compared to 
 $2.5\times$10$^{6}$ red and $6.2\times$10$^{6}$ blue galaxies in the Garching
model. Now the difference between the models in the predicted numbers of red and
blue galaxies is larger than for the case of a fixed color cut. This suggests that
the redshift-dependent color cut selects very different objects in the two models.
Surprisingly, the aperture statistics predicted by the two models are much more
similar for the redshift-dependent color cuts than for the fixed color cut, as seen
in the middle column of Fig. \ref{fig:guo_durham_NN_NM_NNM_paper_red_blue}. The
better agreement results from a decrease in the blue signals in the Durham model
and an increase in the red signals in the Garching model.

The agreement between the two models shown in the middle column of Fig.
\ref{fig:guo_durham_NN_NM_NNM_paper_red_blue} indicates that although the redshift
distributions of red and blue galaxies differ, galaxies populate dark
matter halos in such a way to produce similar results. This agreement between the models is more prominently seen in $\bev{\Nap^2 }$ and $\bev{\Nap^2 \Map}$. 
Looking at $\bev{\Nap \Map}$, red Durham
galaxies show a stronger signal on intermediate scales compared to the red  Garching
galaxies. This can happen, for example, if red galaxies in the Durham model are
mostly central galaxies populating large massive halos. On the other hand, this difference may also be a result of the distinct redshift distributions. As will discussed this difference is not seen when galaxies are restricted to come from a single redshift.

Selecting lens galaxies at a single redshift amplifies the signal for $\bev{\Nap^2}$
and $\bev{\Nap^2 \Map}$ compared to a sample with a broad redshift distribution,
where many of the projected galaxy pairs are at different redshifts and are
therefore not correlated and suppress the overall signal. The third column of
Fig. \ref{fig:guo_durham_NN_NM_NNM_paper_red_blue} displays the aperture
measurements for lens galaxy populations selected from a single redshift slice
around $z = 0.17$ with thickness $\Delta z=0.02$. Now the signals for red galaxies agree
well between the two models. The agreement is not so good for blue model galaxies,
where the Garching model shows stronger signals on small and intermediate scales. In
the halo-model language, blue Garching model galaxies at this redshift appear to
live in more massive halos.

Both the Durham and the Garching models predict a larger ratio between the
clustering strength of red and blue galaxies than has been obtained in observations.
A similar behavior was seen when considering the previous incarnation of the
Garching model based on \citet{2007MNRAS.375....2D}. In particular, our results
confirm the previous work of \cite{2011A&A...525A.125D}, who compared the
color-dependent projected two-point correlation function of a color subsample of
galaxies in the VIMOS-VLT Deep Survey (VVDS; \citealt{2005A&A...439..845L}) and in
the model based on \citet{2007MNRAS.375....2D}. They showed that red galaxies in the
semi-analytic models have stronger clustering amplitudes than the observed ones.
They linked this discrepancy to an overproduction of bright red galaxies in the
model. 
 
The different clustering strengths of red and blue galaxies show up very clearly in
$\bev{\Nap^2}$, $\bev{\Nap\Map}$, and $\bev{\Nap^2\Map}$. The ratio of the
clustering amplitude of the red and blue samples is related to their relative bias
(Sect. \ref{sec:galaxy_bias}). This ratio can be measured based on different
aperture statistics measurements presented in Fig.
\ref{fig:guo_durham_NN_NM_NNM_paper_red_blue}. Assuming a simple linear
deterministic bias, the relative bias and its uncertainty is calculated on aperture
scales of $\theta\sim1 \,\arcmint$ and $\theta\sim10 \,\arcmint$ in the Garching and
Durham models. The results are shown in Tab. \ref{tab:relative-bias}.

\begin{table}[htbp]
\caption{The relative bias $b_\text{red} / b_\text{blue}$ based on different
aperture statistics (Sect. \ref{sec:galaxy_bias}) measured according to the right
column of Fig. \ref{fig:guo_durham_NN_NM_NNM_paper_red_blue} on scales of 1 and 10
$\arcmint$ in the Garching and Durham models. The values are obtained
assuming a linear deterministic bias model. }
\begin{tabular}{ccccc}
&  & $\bev{\Nap^2}$ & $\bev{\Nap\Map}$ & $\bev{\Nap^2\Map}$   \\ \cline{1-5}
\multicolumn{1}{|c|}{\multirow{2}{*}{1$^{\prime}$}} &
\multicolumn{1}{|c|}{Garching} & 5.4$\pm$0.1 & 6.06$\pm$0.21 &
\multicolumn{1}{c|}{10.83$\pm$0.32}    \\ \cline{2-5}
\multicolumn{1}{|c|}{}                        &
\multicolumn{1}{|c|}{Durham} & 9.73$\pm$0.13 & 9.34$\pm$0.37 &
\multicolumn{1}{r|}{18.9$\pm$0.64}    \\ \cline{1-5}
\multicolumn{1}{|c|}{\multirow{2}{*}{10$^{\prime}$}} &
\multicolumn{1}{|c|}{Garching} & 3.37$\pm$0.19  & 3.84$\pm$0.93 &
\multicolumn{1}{r|}{4.8$\pm$0.94} \\ \cline{2-5}
\multicolumn{1}{|c|}{}                        &
\multicolumn{1}{|c|}{Durham} & 3.52$\pm$0.21 & 4.82$\pm$1.1 &
\multicolumn{1}{r|}{7.15$\pm$1.5}   \\ \cline{1-5}
\end{tabular}
\label{tab:relative-bias}
\end{table}

The differences in the bias ratios measured from different statistics point out that
a linear deterministic bias model is not sufficient to describe the relation between
the galaxy and matter distribution on different scales. This relation may be
described by scale-dependent stochastic bias.

\subsection{Magnitude-selected samples}

\begin{figure*}
\centerline{\includegraphics[width=120mm]{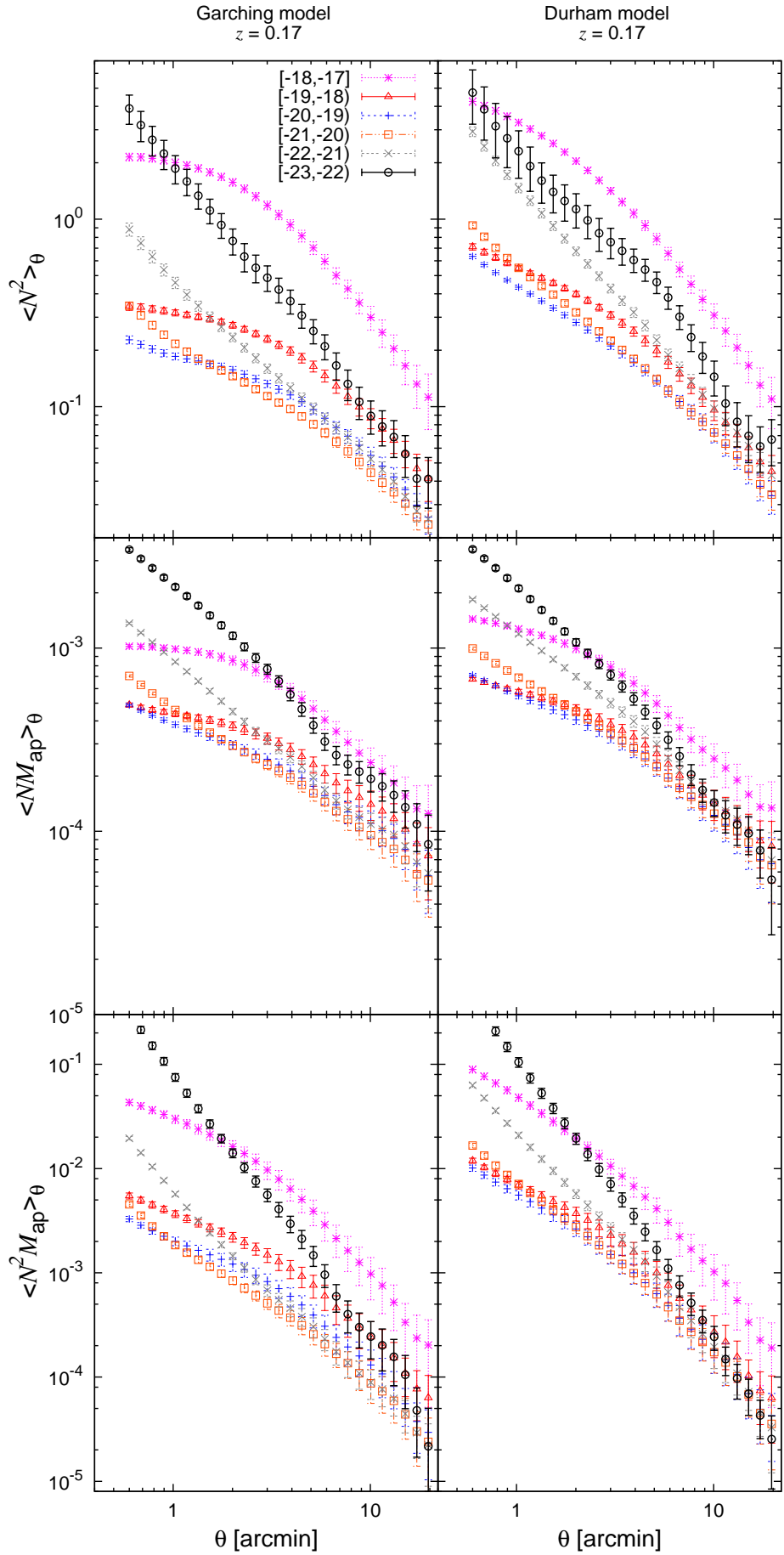}}
\caption{
Aperture measurements in the Garching model (left panel) and the Durham model (right
panel) in 6 different $r$-band absolute magnitude, $M_r$, bins.}
\label{fig:guo_durham_NN_NM_NNM_magnitude}
\end{figure*}

In this section, we present the measurements of the second- and third-order aperture
statistics for lens galaxies in six different bins of $r$-band absolute observer
frame magnitude $M_{r}$. To eliminate effects of possibly different redshift
distributions on the signals, we restrict the redshift range of the lens galaxy
population to one redshift slice at $z = 0.17$. The results for all magnitude bins
for the Garching (Durham) model are shown in the left (right) panels of Fig.
\ref{fig:guo_durham_NN_NM_NNM_magnitude}.

There are common trends seen for both models in the second- and third-order aperture
statistics. For the brighter bins ($-23 \leq M_r < -20$), the aperture signals
decrease rapidly with increasing magnitude $M_r$ and filter scales $\theta$.
However, the Durham model predicts up to 200\% higher $\bev{\Nap^2\Map}$, and
$\bev{\Nap^2}$ signals than the Garching model. Bright galaxies appear more
clustered and on average to be located in more massive halos in the Durham model.

For the fainter bins ($-20 \leq M_r < -18$), the signals \emph{increase} with
decreasing luminosity. This increase is contrary to observations of galaxy
clustering and GGL (see e.g. \citealt{2011ApJ...726...13M}), where brighter galaxies
show stronger clustering and larger lensing signals than fainter galaxies.

In the Garching model, the faint magnitude bins are over-populated with
satellite galaxies, many of which have no own subhalo (this occurs when a galaxy has
been stripped of its own halo during a merger process with a larger halo). These
galaxies are abundant in massive halos, which contribute substantially to the
$\bev{\Nap\Map}$ and $\bev{\Nap^2\Map}$ signal due to their large mass and stronger clustering. In
the Durham model a similar trend is seen in $\bev{\Nap\Map}$ and $\bev{\Nap^2\Map}$,
indicating similar problems with the modeling of the satellite population in
massive halos.

The luminosity dependence of galaxy clustering has been studied extensively with the
aid of galaxy surveys. \cite{2007MNRAS.376..984L} compared the luminosity dependence
of the clustering of galaxies in the model of \cite{2006MNRAS.365...11C} to results
from the Sloan Digital Sky Survey Data Release Four \citep[SDSS
DR4;][]{2006ApJS..162...38A}, and found that the faint model galaxies show a
stronger clustering than SDSS galaxies. 

\cite{2009MNRAS.400.1527K} compared the galaxy clustering predicted by the models of
\citet{2006MNRAS.370..645B}, \citet{2007MNRAS.375....2D} and
\citet{2008MNRAS.389.1619F} to observed clustering in the two-degree Field Galaxy
Redshift Survey \citep[2dF,][]{2001MNRAS.328.1039C}. They found that none of the
models are able to match the observed clustering properties of galaxies in different
luminosity bins. In particular, the Durham model shows a stronger signal than
expected, which could possibly be corrected, if the number of satellite galaxies in
halos is reduced.

Both \cite{2007MNRAS.376..984L} and \cite{2009MNRAS.400.1527K} emphasize
the problems of the galaxy models in predicting the luminosity dependence of galaxy
clustering. \cite{2007MNRAS.376..984L} showed that the number of faint satellite
galaxies has to be reduced by 30 per cent (regardless of their host halo mass) to
better match the observed galaxy clustering. \cite{2009MNRAS.400.1527K} showed that
the fraction of satellites declines with increasing luminosity \citep[see, e.g. Fig
4 in][]{2009MNRAS.400.1527K} in the host halo mass range of $10^{12}h^{-1}\,\Msolar
\lesssim M_{\textrm{halo}} \lesssim 10^{14}h^{-1}\,\Msolar$. Since the clustering
strength depends strongly on the halo mass, satellite galaxies can affect the
overall clustering amplitude. To investigate this they used a simple HOD model to
show that satellite galaxies show a strong bias due to a strong two-halo clustering
term. This indicates that satellite galaxies are preferentially found in massive halos
which exhibits larger bias \citep[see Fig. 5 in][]{2009MNRAS.400.1527K}. They argued
that the results can be improved if the satellites are removed from massive halos by adding satellite-satellite merger processes in the models.

Our results suggest that the Garching model shows a similar problem with faint
satellite galaxies, though to a lesser degree.
Indeed, we find that the amplitude of the aperture signals in the faint bins are
completely dominated by satellite galaxies in both the Durham and the Garching
model. Most of these faint satellites reside in massive group and cluster halos,
which results in very large aperture signals on the scales considered in this work.

\section{Summary and discussion}
\label{sect:discussion}

Through observations of galaxy-galaxy(-galaxy) lensing, valuable information on the
clustering properties of the galaxy and matter density field in the Universe can be
obtained. Measurements of galaxy-galaxy lensing (GGL) can be used to infer
information on the properties of dark matter halos hosting the lens galaxies
\citep[see, e.g.,][]{1997ApJ...474...25S,2007arXiv0709.1159J,2008JCAP...08..006M}.
Third-order galaxy lensing (G3L) can be used to infer information on the properties
of a common dark matter halo hosting two lens galaxies
\citep[]{2008A&A...479..655S,2012arXiv1202.1927S}.

In this work, we study how the information from GGL and G3L aperture statistics can
be used to test models of galaxy formation and evolution. 
We investigate two semi-analytic galaxy formation models based on the Millennium Run
N-body simulation of structure formation \citep{2005Natur.435..629S}: the Durham
model by \citet{2006MNRAS.370..645B}, and the Garching model by
\citet{2011MNRAS.413..101G}. Using mock galaxy catalogs based on these models in
conjunction with ray-tracing \citep{2009A&A...499...31H}, we create simulated fields
of galaxy lensing surveys. From these simulated surveys, we compute the model
predictions for the second- and third-order aperture statistics $\bev{\Nap^2}$,
$\bev{\Nap \Map}$, and $\bev{\Nap^2 \Map}$ for various galaxy populations. 

We find that both semi-analytic models predict aperture signals that are
qualitatively similar, but there are large quantitative differences. The Durham
model predicts larger amplitudes for most considered galaxy samples. This indicates
that lens galaxies in the Durham model tend to reside in more massive halos than
lens galaxies in the Garching model.

In both the Durham and the Garching model, red galaxies exhibit stronger aperture
signals than blue galaxies, in qualitative agreement with observations. However,
both models predict a larger ratio between the clustering strength of red and blue
galaxies than has been obtained in observations. These findings  corroborate the
findings of \citet{2011A&A...525A.125D}, who showed that red galaxies in the
semi-analytic models have stronger clustering amplitudes than red galaxies in
observations.
 We argue that considering the amplitude ratio between the red
and blue galaxies and making comparison between the second- and third-order aperture
statistics leads to the conclusion that the third-order bias differs from the
second-order bias. In other words, third-order aperture statistics provides new
information which cannot be obtained from the second-order statistics alone.
The large amplitude ratio between the clustering of red and blue galaxies in the
models means a large relative bias of these galaxy populations. Measuring the
biasing of galaxies provides information on the relative distribution of galaxies
and the underlying matter distribution.  We find that a linear deterministic bias
model, even with
scale-dependent bias parameters, is clearly ruled out by considering
second and third-order aperture statistics for the simulated data. We
expect that both statistics in combination will provide new information
to constrain more advanced galaxy biasing models in the future.

In addition to the different prediction for red and blue galaxies, there are
discrepancies between the predictions of the two models. For a fixed color cut at $u
- r = 2.2$, the signals predicted by the Durham model are larger than those
predicted by the Garching model. If a redshift-dependent color cut is used instead,
the prediction from the two models for the aperture signals become more similar.
However, the models then strongly disagree about the total numbers and redshift
distributions of blue and red galaxies.

Both galaxy models predict that the aperture statistics decrease with decreasing
luminosity for brighter galaxies in accordance with observations. However, the
models also predict that the signals increase again for fainter galaxies. This
behavior is most likely an artifact related to too many faint satellite galaxies in
massive group and cluster halos predicted by the models. In fact, the fainter
magnitude bins are completely dominated by satellite galaxies in both models. The
problem appears more severe in the Durham model than in the Garching model, which
differ in their treatment of satellite evolution.

We plan to extent our treatment in future work to study how well galaxy bias models
with scale-dependent stochastic bias can be constrained with second- and third-order
galaxy lensing statistics. One important question is how much information
can be obtained from G3L in addition to that obtained from GGL.

Furthermore, we are looking forward to measurements of GGL and G3L signals in large
ongoing and future surveys. The comparison of the observed signals and the signals
predicted by galaxy models will help to identify shortcomings of the models and
provide valuable hints for improvements in the models. This will also require a
deeper understanding of the relation between the various details of the galaxy
formation models and the predicted galaxy lensing and clustering signals.

\acknowledgements
We thank Phil Bett and Peder Norberg for helpful discussions.
We thank the people of the Virgo Consortium for Cosmological Supercomputer
Simulations and the German Astrophysical Virtual Observatory and all other people
involved in making the Millennium Run simulation data and the galaxy catalogs of the
semi-analytic galaxy formation models used in this work publicly available.  
This work was supported by the Deutsche Forschungsgemeinschaft (DFG) through the
Priority Programme 1177 `Galaxy Evolution' (SCHN 342/6, SCHN 342/8--1, and WH 6/3)
and through the Transregional Collaborative Research Centre TRR 33 `The Dark
Universe'.
SH also acknowledges support by the National Science Foundation (NSF) grant number
AST-0807458-002.

\appendix

\section{Shot-noise Correction}
\label{sec:shot_noise_correction}

Assume a realization $\bigl(\vvartheta^{(r)}_{i}\bigr)$, $i = 1,\ldots,\Ngal$, of a
set of $\Ngal$ galaxies with sky positions $\vvartheta^{(r)}_{i}$ in a field
$\mathcal{A}$ with area $A$ and mean galaxy number density of $\meanngal = \Ngal /
A$. Assume that each galaxy position is distributed in the field according to an
underlying `true' number density field $\ngal(\vvartheta)$. The ensemble average of
a quantity $o(\vvartheta)$ over all realizations reads:
\begin{equation}
   \left\langle o(\vvartheta) \right\rangle  =  \left[ \prod_{k=1}^{\Ngal}
\frac{1}{\Ngal} \int_{\mathcal{A}}\dd^2 \vvartheta^{(r)}_{k} \,
n(\vvartheta^{(r)}_{k}) \right] o(\vvartheta)\, .
\end{equation}

For each realization, the random positions of the galaxies provide an estimate of
the density field $\ngal$:
\begin{equation}
\label{eq:ngal_estimator}
    \estngal^{(r)}(\vvartheta) = \sum_{i=1}^{\Ngal} \deltaDirac \left(\vvartheta -
\vvartheta^{(r)}_{i} \right).
\end{equation}
This estimator is unbiased:
\begin{equation}
\begin{split}
\ev{ \estngal^{(r)}(\vvartheta) } &=
  \left[ \prod_{k=1}^{\Ngal} \frac{1}{\Ngal} \int_{\mathcal{A}}\dd^2
\vvartheta^{(r)}_{k} \,  \ngal(\vvartheta^{(r)}_{k}) \right]
  \sum_{i=1}^{\Ngal} \deltaDirac \left(\vvartheta - \vvartheta^{(r)}_{i} \right)
\\&=
  \sum_{i=1}^{\Ngal}
  \left[ \prod_{k=1}^{\Ngal} \frac{1}{\Ngal} \int_{\mathcal{A}}\dd^2
\vvartheta^{(r)}_{k} \,  \ngal(\vvartheta^{(r)}_{k}) \right]
  \deltaDirac \left(\vvartheta - \vvartheta^{(r)}_{i} \right)
\\&=
  \sum_{i=1}^{\Ngal}
   \frac{1}{\Ngal} \int_{\mathcal{A}}\dd^2 \vvartheta^{(r)}_{i} \, 
\ngal(\vvartheta^{(r)}_{i})
  \deltaDirac \left(\vvartheta - \vvartheta^{(r)}_{i} \right)
\\&= 
   \ngal(\vvartheta)
.
\end{split}
\end{equation}

Using a filter function $U(\vvartheta)$, we define a filtered density field $\Nap$ by:
\begin{equation}
    \Nap(\vvartheta; U) = \int_{\mathcal{A}}\dd^2 \vvartheta'\, U(\vvartheta -
\vvartheta')\,\ngal(\vvartheta')\,.
\end{equation}
An estimator for the filtered field reads:
\begin{equation}
\label{eq:nap_estimator}
    \est{\Nap}^{(r)}(\vvartheta; U) =
     \int_{\mathcal{A}}\dd^2 \vvartheta'\, U(\vvartheta -
\vvartheta')\,\estngal^{(r)}(\vvartheta')
     =
     \sum_{i=1}^{\Ngal} U(\vvartheta - \vvartheta^{(r)}_{i})
    .
\end{equation}
Its expectation value 
\begin{equation}\begin{split}
  \left\langle \est{\Nap}^{(r)}(\vvartheta; U) \right\rangle_{r} &= \left [
\prod_{k=1}^{\Ngal} \frac{1}{\Ngal} \int_{\mathcal{A}}\dd^2 \vvartheta^{(r)}_{k}
\,  \ngal(\vvartheta^{(r)}_{k}) \right ]\sum_{i=1}^{\Ngal} U(\vvartheta -
\vvartheta^{(r)}_{i})\\
    &= \sum_{i=1}^{\Ngal}\frac{1}{\Ngal} \int_{\mathcal{A}}\dd^2
\vvartheta^{(r)}_{i} \,  \ngal(\vvartheta^{(r)}_{i}) U(\vvartheta -
\vvartheta^{(r)}_{i})\\
    &= \int_{\mathcal{A}}\dd^2 \vvartheta'\, \ngal(\vvartheta') U(\vvartheta -
\vvartheta')\\
    &= \Nap(\vvartheta; U)
.
\end{split}\end{equation}

Consider the square of the filtered density
\begin{equation}
    \Nap^{2}(\vvartheta; U) = 
     \int_{\mathcal{A}}\dd^2 \vvartheta'\int_{\mathcal{A}}\dd^2 \vvartheta'' \, 
U(\vvartheta - \vvartheta') \,
     U(\vvartheta - \vvartheta'')\,\ngal(\vvartheta')\,\ngal(\vvartheta'')
    .
\end{equation}
A naive estimator is provided by:
\begin{equation}
\begin{split}
\ev{\bigl[\est{\Nap}^{(r)}(\vvartheta; U)\bigr]^{2}} &=
  \left[ \prod_{k=1}^{\Ngal} \frac{1}{\Ngal} \int_{\mathcal{A}}\dd^2
\vvartheta^{(r)}_{k}\, \ngal(\vvartheta^{(r)}_{k}) \right]
  \\&\quad\qquad\times
  \,\sum_{i=1}^{\Ngal} U(\vvartheta - \vvartheta^{(r)}_{i}) \sum_{j=1}^{\Ngal}
U(\vvartheta - \vvartheta^{(r)}_{j})
  \!\!\!\!
\\&=
  \sum_{\substack{i\neq j\\i,j=1 }}^{\Ngal}\frac{1}{\Ngal^{2}}
  \int_{\mathcal{A}}\dd^2 \vvartheta^{(r)}_{i}\int_{\mathcal{A}}\dd^2
\vvartheta^{(r)}_{j}\, \ngal(\vvartheta^{(r)}_{i})\, \ngal(\vvartheta^{(r)}_{j})
  \\&\quad\qquad\times
  \,U(\vvartheta - \vvartheta^{(r)}_{i})\,U(\vvartheta - \vvartheta^{(r)}_{j})
  \\&\quad +
  \sum_{i=1}^{\Ngal}\frac{1}{\Ngal} \int_{\mathcal{A}}\dd^2 \vvartheta^{(r)}_{i} \,
\ngal(\vvartheta^{(r)}_{i}) U(\vvartheta - \vvartheta^{(r)}_{i})^{2}
\\&= 
   \frac{\Ngal(\Ngal - 1)}{\Ngal^{2}} \Nap^{2}(\vvartheta;U)
   +  \Nap(\vvartheta; U^2)
. 
\label{eq:shot-noise}
\end{split}
\end{equation}
Hence, this estimator is biased. The first term of the last line is actually what is
intended to be measured as aperture dispersion (up to a prefactor close to unity).
The second term is due to shot noise arising from the Poisson sampling of the
density field. An unbiased estimator of $\Nap^2$ is provided by 
\begin{equation}
\begin{split}
\label{eq:nap2_estimator}
&\frac{\Ngal}{\Ngal - 1}
\left\{
\bigl[\est{\Nap}^{(r)}(\vvartheta;U)\bigr]^{2} - \est{\Nap}^{(r)}(\vvartheta;U^2)
\right\}
=
\frac{\Ngal}{\Ngal - 1}
\\&\quad\qquad\times
\sum_{\substack{i\neq j\\i,j=1 }}^{\Ngal}
U(\vvartheta - \vvartheta^{(r)}_{i}) U(\vvartheta - \vvartheta^{(r)}_{j})
.
\end{split}
\end{equation}

By projecting the galaxy positions of a realization onto a mesh (e.g. using
Nearest-Grind-Point assignment), one obtains a discretized representation of the
density estimate \eqref{eq:ngal_estimator}. The density estimate on the mesh can
then be convolved, e.g. by using FFTs, with the filters $U$ and $U^2$ to obtain 
gridded versions of the unbiased estimates for $\Nap_{U}$ and $\Nap_{U^2}$. The
latter estimate can then be subtracted point-wise from the square of the former
estimate to calculate the unbiased estimate \eqref{eq:nap2_estimator}.

\bibliographystyle{aa}
\bibliography{BibFiles}

\end{document}